\documentclass[conference]{IEEEtran}
\usepackage[a4paper, left=0.5in, right=0.5in, bottom=1.5in, top=0.75in]{geometry}

\usepackage{booktabs} 

\usepackage[utf8]{inputenc} 
\usepackage[T1]{fontenc}
\usepackage{url}
\usepackage{cite}
\usepackage{amsmath} 
\usepackage{amssymb}  
\usepackage{amsthm}
\usepackage{enumerate,color,float}
\usepackage[justification=centering]{caption}
\usepackage{amsfonts}
\usepackage{mathtools}
\usepackage{arydshln}
\usepackage{stmaryrd}
\usepackage{cases}
\usepackage{multicol}
\usepackage{algorithm}
\usepackage{algorithmic}
\usepackage{tikz,pgfplots,rotating}
\usepackage{bm}
\DeclareMathAlphabet{\mathcal}{OMS}{cmsy}{m}{n} 
\usetikzlibrary{shapes, positioning, arrows,patterns}
\usepackage{tikz,amsmath, amssymb,bm,color} 
\usetikzlibrary{circuits.logic.US,circuits.logic.IEC,fit} 
\newtheoremstyle{remboldstyle}
  {}{}{}{}{\bfseries}{.}{.5em}{{\thmname{#1 }}{\thmnumber{#2}}}
\theoremstyle{plain}
\newtheorem{theorem}{Theorem}

\theoremstyle{definition}

\newtheorem{example}[theorem]{Example}

\theoremstyle{remboldstyle}

\theoremstyle{corollary}

\theoremstyle{lemma}

\newcommand{\mb}[1]{\mathbf{#1}}

\newcommand{\mbb}[1]{\mathbb{#1}}
\newcommand{\GF}[1]{\mbb{F}_{#1}}

\newcommand{\Rm}[1]{\mathrm{#1}}

\newcommand{\boldV}[1]{\boldsymbol{\mathbf{#1}}}

\newcommand{\om}{\omega}               

\newcommand{\defeq}{\triangleq}

\newcommand{\fo}[1]{\makebox[1ex][c]{$0$}}
\newcommand{\fl}[1]{\makebox[1ex][c]{$1$}}
\newcommand{\fw}[1]{\makebox[1ex][c]{$\om$}}
\newcommand{\fwb}[1]{\makebox[1ex][c]{$\bar{\om}$}}

\newcommand{\vf}{\mathbf{f}}

\newcommand{\vs}{\mathbf{s}}

\newcommand{\vv}{\mathbf{v}}

\newcommand{\tr}{\mathsf{T}}

\newcommand{\Th}{\Rm{th}}
\DeclareMathOperator*{\argmin}{\arg\,\min}

\DeclareMathOperator{\Supp}{supp}

\newcommand{\qedblack}{\hspace*{\fill}%
  $\blacksquare$\smallskip}

\newcommand{\exampleend}{\qedblack}

\newcommand{\cB}{\mathcal{B}}
\newcommand{\ctB}{\tilde{\mathcal{B}}}
\newcommand{\cC}{\mathcal{C}}

\newcommand{\cG}{\mathcal{G}}

\newcommand{\cJ}{\mathcal{J}}

\newcommand{\CASE}[1]{\STATE \textbf{case} #1\textbf{:} \begin{ALC@g}}
\newcommand{\ENDCASE}{\end{ALC@g}}

\newcommand{\DEFAULT}{\STATE \textbf{default:} \begin{ALC@g}}
\newcommand{\ENDDEFAULT}{\end{ALC@g}}
\newcommand{\DEFAULTLINE}[1]{\STATE \textbf{default:} }

\newlength{\myl}

\newcommand{\Footnotetext}[2]{\begin{figure}[!b]\footnotesize
  \vspace{-3ex}\hrulefill\hfill\makebox[0em]{}\hfill\makebox[0em]{}
  \par${}^{#1}$ #2\vspace{-0.60ex}\end{figure}\addtocounter{figure}{0}}

\begin{document}
\title{Pseudocodeword-based Decoding \\of Quantum Color Codes}

\author{%

  \IEEEauthorblockN{July X.\ Li\IEEEauthorrefmark{1}, 
  						Joseph M.\ Renes\IEEEauthorrefmark{2}, and 
   						Pascal O.\ Vontobel\IEEEauthorrefmark{3}}
  \IEEEauthorblockA{\IEEEauthorrefmark{1}\IEEEauthorrefmark{3}Department of Information Engineering,
                    The Chinese University of Hong Kong, 
                    Hong Kong\\
\IEEEauthorrefmark{2}Institute for Theoretical Physics, ETH Zürich, 8093 Zürich, Switzerland\\
                    Email: \IEEEauthorrefmark{1}\IEEEauthorrefmark{3}\{july.x.li, 
                    		  pascal.vontobel\}@ieee.org, 
                    		  \IEEEauthorrefmark{2}renes@phys.ethz.ch\vspace{-0.5cm}}
}

\maketitle


\begin{abstract}
In previous work, we have shown that pseudocodewords can be used to characterize the behavior of decoders not only for classical codes but also for quantum stabilizer codes. 
With the insights obtained from this pseudocodewords-based analysis, we have also introduced a two-stage decoder based on pseudocodewords for quantum cycle codes that leads to improved decoding performance.
In this paper, we consider quantum (stabilizer) color codes and propose a two-stage decoder that is a generalization of the  pseudocodeword-based decoder for quantum cycle codes.
Our decoder has only local or error-weight-dependent operations of low computational complexity and better decoding performance compared with previous decoding approaches for these types of codes.
\end{abstract}

\Footnotetext{}{This work was supported by the Research Grants Council of the Hong Kong Special Administrative Region, China, under Project CUHK 14209317 and Project CUHK 14208319,
the Swiss National Science Foundation (SNSF), National Center of Competence in Research "QSIT", and 
Air Force Office of Scientific Research (AFOSR), grant FA9550-19-1-0202.
}
\section{Introduction}
\label{section1:introduction}

Graph covers and pseudocodewords have been shown to be a useful tool for analyzing sum-product algorithm (SPA) decoding of classical codes~\cite{vontobel2013counting}. 
Although the task of analyzing the behavior of decoders for quantum stabilizer codes is more challenging because of the degeneracy of quantum stabilizer codes, it is highly desirable to understand and improve decoders also for such codes.
We have initiated the use of pseudocodewords to analyze and characterize decoders for stabilizer codes, including LP decoding~\cite{li2018lp} and SPA decoding~\cite{li2019pseudocodeword}.
Based on the insights obtained using pseudocodewords to analyze the behavior of SPA decoding, we have introduced a two-stage pseudocodeword-based decoder for quantum cycle codes in~\cite{li2019pseudocodeword} working as follows. 
1) Run the SPA. 
2) If the SPA outputs a pseudocodeword that is a codeword then stop. 
3) If the SPA outputs a non-codeword pseudocodeword then decompose the pseudocodeword to get paths and then use a linear program (LP) to choose a collection of paths corresponding to a vector with a matching syndrome.

In this paper, we further generalize the two-stage pseudo\-codeword-based decoder for quantum cycle codes toward decoding quantum (stabilizer) color codes.
Note that it is much more difficult to decode quantum color codes compared with quantum cycle codes since the variable node degree increases from $2$ to $3$ when going from quantum cycle codes to quantum color codes.
For decoding a quantum color code, we first project a quantum color code into cycle codes as in~\cite{delfosse2014decoding,kubica2019efficient}, 
then convert paths obtained from the two-stage pseudocodeword-based decoder for quantum cycle codes back to ``generalized paths'', 
and finally use an error-weight-dependent LP to choose some ``generalized paths'' to get a vector with a matching syndrome.
Problems of previous approaches~\cite{delfosse2014decoding,kubica2019efficient} include high computational complexity and no guarantee of obtaining a vector with a matching syndrome.
Our decoder is special in the following sense. 
First, we use a pseudocodeword-based decoder which makes the best use  of the pseudocodewords found using SPA decoding for cycle codes instead of using the minimum-weight perfect matching (MWPM) algorithm for quantum cycle codes.
Second, all the operations are local and of low computational complexity compared with~\cite{delfosse2014decoding}.
Finally, no path is dropped before making the final decision using the LP, which can avoid some decoding failures that happen when using the methods in~\cite{delfosse2014decoding,kubica2019efficient}.

This paper is organized as follows.
In Section~\ref{section2:color code}, we review some basic notations and definitions for quantum color codes.
In Section~\ref{section3:PCWDdecoding}, we propose a two-stage pseudocodeword-based decoder for quantum color codes.
Finally, we show some simulation results in Section~\ref{section4:simulation}.

\section{Quantum Color Codes}
\label{section2:color code}
\begin{figure}
\begin{center}
\resizebox{0.6\columnwidth}{!}{\begin{tikzpicture}[x=1cm, y= 1cm, node/.style={draw=none},
factor/.style={draw,circle,minimum size = 9pt},
factorR/.style={draw,circle,minimum size = 9pt,color=red},
factorB/.style={draw,circle,minimum size = 9pt,color=blue},
factorG/.style={draw,circle,minimum size = 9pt,color=green},
factorRg/.style={draw,circle,minimum size = 9pt,color=red!20},
factorBg/.style={draw,circle,minimum size = 9pt,color=blue!20},
factorGg/.style={draw,circle,minimum size = 9pt,color=green!20}
]
                          
\foreach \i in {0,1,2,3,4,5}{
\foreach \j in {0,1,2,3,4,5}{
\pgfmathparse{int(mod(\i+\j,3))} \let\tf\pgfmathresult 
\ifthenelse{\tf=0}{
	\node[factorR] (node\i\j) at (\i,-\j) {};
	\node at (node\i\j) {};
}{}; 
\ifthenelse{\tf=1}{
	\node[factorB] (node\i\j) at (\i,-\j) {};
	\node at (node\i\j) {};
}{}; 
\ifthenelse{\tf=2}{
	\node[factorG] (node\i\j) at (\i,-\j) {};
	\node at (node\i\j) {};
}{};  
}}      
\foreach \i in {0,1,2,3,4,5}{
\foreach \j in {6}{
\pgfmathparse{int(mod(\i+\j,3))} \let\tf\pgfmathresult 
\ifthenelse{\tf=0}{
	\node[factorRg] (node\i\j) at (\i,-\j) {};
	\node at (node\i\j) {};
}{}; 
\ifthenelse{\tf=1}{
	\node[factorBg] (node\i\j) at (\i,-\j) {};
	\node at (node\i\j) {};
}{}; 
\ifthenelse{\tf=2}{
	\node[factorGg] (node\i\j) at (\i,-\j) {};
	\node at (node\i\j) {};
}{};  
}}   
\foreach \i in {6}{
\foreach \j in {0,1,2,3,4,5,6}{
\pgfmathparse{int(mod(\i+\j,3))} \let\tf\pgfmathresult 
\ifthenelse{\tf=0}{
	\node[factorRg] (node\i\j) at (\i,-\j) {};
	\node at (node\i\j) {};
}{}; 
\ifthenelse{\tf=1}{
	\node[factorBg] (node\i\j) at (\i,-\j) {};
	\node at (node\i\j) {};
}{}; 
\ifthenelse{\tf=2}{
	\node[factorGg] (node\i\j) at (\i,-\j) {};
	\node at (node\i\j) {};
}{};  
}}                       
       
\foreach \i in {0,1,2,3,4,5}{
\foreach \j in {0,1,2,3,4,5}{
\pgfmathparse{int(mod(\i+\j,3))} \let\tf\pgfmathresult 
\pgfmathtruncatemacro\jj{\j+1}; 
\pgfmathtruncatemacro\ii{\i+1}; 
\ifthenelse{\tf=0}{
		\draw[green] (node\i\j) -- (node\i\jj); 
		\draw[green] (node\i\j) -- (node\ii\j);
		\draw[blue] (node\i\j) -- (node\ii\jj); 
}{}
\ifthenelse{\tf=1}{
		\draw[red] (node\i\j) -- (node\i\jj); 
		\draw[red] (node\i\j) -- (node\ii\j);
		\draw[green] (node\i\j) -- (node\ii\jj); 
}{}
\ifthenelse{\tf=2}{
		\draw[blue] (node\i\j) -- (node\i\jj); 
		\draw[blue] (node\i\j) -- (node\ii\j);
		\draw[red] (node\i\j) -- (node\ii\jj); 
}{}
}}     
              
\foreach \i in {6}{
\foreach \j in {0,1,2,3,4,5}{   
\pgfmathparse{int(mod(\i+\j,3))} \let\tf\pgfmathresult 
\pgfmathtruncatemacro\jj{\j+1}; 
\ifthenelse{\tf=0}{
		\draw[green] (node\i\j) -- (node\i\jj); 
}{}
\ifthenelse{\tf=1}{
		\draw[red] (node\i\j) -- (node\i\jj); 
}{}
\ifthenelse{\tf=2}{
		\draw[blue] (node\i\j) -- (node\i\jj); 
}{}
}}    

\foreach \i in {0,1,2,3,4,5}{
\foreach \j in {6}{
\pgfmathparse{int(mod(\i+\j,3))} \let\tf\pgfmathresult 
\pgfmathtruncatemacro\ii{\i+1}; 
\ifthenelse{\tf=0}{
		\draw[green] (node\i\j) -- (node\ii\j);
}{}
\ifthenelse{\tf=1}{
		\draw[red] (node\i\j) -- (node\ii\j);
}{}
\ifthenelse{\tf=2}{
		\draw[blue] (node\i\j) -- (node\ii\j);
}{}
}}         

\end{tikzpicture}}
\end{center}
\caption{Graph $\cG$ used to define the $\llbracket 72, 4, 8\rrbracket$ color code $\cC$, 
where the vertices (in lighter color) in the last row/column are repeated from the vertices in the first row/column.}
\label{fig:color_C_X_dual2}
\end{figure}
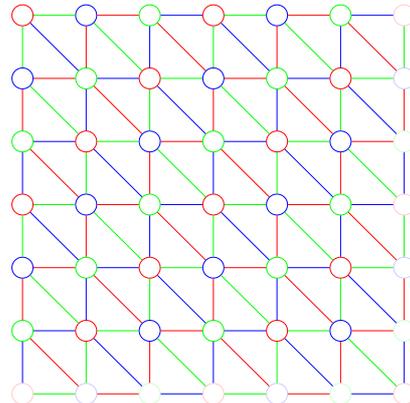

\begin{figure}
\begin{center}
\resizebox{0.6\columnwidth}{!}{\begin{tikzpicture}[x=1cm, y= 1cm, node/.style={draw=none},
factor/.style={draw,circle,minimum size = 9pt},
factorR/.style={draw,circle,minimum size = 9pt,color=red},
factorB/.style={draw,circle,minimum size = 9pt,color=blue},
factorG/.style={draw,circle,minimum size = 9pt,color=green},
factorRg/.style={draw,circle,minimum size = 9pt,color=red!20},
factorBg/.style={draw,circle,minimum size = 9pt,color=blue!20},
factorGg/.style={draw,circle,minimum size = 9pt,color=green!20}
]
                          
\foreach \i in {0,1,2,3,4,5}{
\foreach \j in {0,1,2,3,4,5}{
\pgfmathparse{int(mod(\i+\j,3))} \let\tf\pgfmathresult 
\ifthenelse{\tf=0}{}{}; 
\ifthenelse{\tf=1}{
	\node[factorB] (node\i\j) at (\i,-\j) {};
	\node at (node\i\j) {};
}{}; 
\ifthenelse{\tf=2}{
	\node[factorG] (node\i\j) at (\i,-\j) {};
	\node at (node\i\j) {};
}{};  
}}              
\foreach \i in {0,1,2,3,4,5}{
\foreach \j in {6}{
\pgfmathparse{int(mod(\i+\j,3))} \let\tf\pgfmathresult 
\ifthenelse{\tf=0}{}{}; 
\ifthenelse{\tf=1}{
	\node[factorBg] (node\i\j) at (\i,-\j) {};
	\node at (node\i\j) {};
}{}; 
\ifthenelse{\tf=2}{
	\node[factorGg] (node\i\j) at (\i,-\j) {};
	\node at (node\i\j) {};
}{};  
}}            
\foreach \i in {6}{
\foreach \j in {0,1,2,3,4,5,6}{
\pgfmathparse{int(mod(\i+\j,3))} \let\tf\pgfmathresult 
\ifthenelse{\tf=0}{}{}; 
\ifthenelse{\tf=1}{
	\node[factorBg] (node\i\j) at (\i,-\j) {};
	\node at (node\i\j) {};
}{}; 
\ifthenelse{\tf=2}{
	\node[factorGg] (node\i\j) at (\i,-\j) {};
	\node at (node\i\j) {};
}{};  
}}                        
       
\foreach \i in {0,1,2,3,4,5}{
\foreach \j in {0,1,2,3,4,5}{
\pgfmathparse{int(mod(\i+\j,3))} \let\tf\pgfmathresult 
\pgfmathtruncatemacro\jj{\j+1}; 
\pgfmathtruncatemacro\ii{\i+1}; 
\ifthenelse{\tf=1}{
		\draw[red] (node\i\j) -- (node\i\jj); 
		\draw[red] (node\i\j) -- (node\ii\j);
}{}
\ifthenelse{\tf=2}{
		\draw[red] (node\i\j) -- (node\ii\jj); 
}{}
}}     
              
\foreach \i in {6}{
\foreach \j in {0,1,2,3,4,5}{   
\pgfmathparse{int(mod(\i+\j,3))} \let\tf\pgfmathresult 
\pgfmathtruncatemacro\jj{\j+1}; 
\ifthenelse{\tf=1}{
		\draw[red] (node\i\j) -- (node\i\jj); 
}{}
}}    

\foreach \i in {0,1,2,3,4,5}{
\foreach \j in {6}{
\pgfmathparse{int(mod(\i+\j,3))} \let\tf\pgfmathresult 
\pgfmathtruncatemacro\ii{\i+1}; 
\ifthenelse{\tf=1}{
		\draw[red] (node\i\j) -- (node\ii\j);
}{}
}}         

\end{tikzpicture}}
\end{center}
\caption{Graph ${\cG}_{\Rm{R}}$ for ${\cC}_{\Rm{R}}$, which is the subgraph of $\cG$ in Fig.~\ref{fig:color_C_X_dual2} induced by the set of red edges.}
\label{fig:color_C_X_dual_R}
\end{figure}
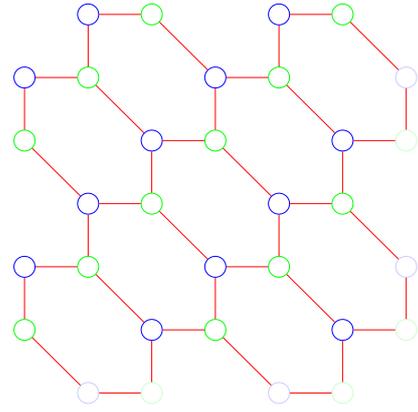

Due to the page limitations, we only introduce essential notations and we refer readers to~\cite{nielsen2010quantum} for a detailed introduction to quantum stabilizer codes.
A Calderbank-Shor-Steane (CSS) code is defined to be a quantum stabilizer code with binary stabilizer label code $\ctB\defeq \cC_X^{\perp} \times \cC_Z^{\perp}$,
where $\cC_X$ and $C_Z$ are two binary linear codes satisfying $\cC_X^{\perp}\subseteq \cC_Z$.
Quantum color codes, henceforth simply called color codes, are CSS codes defined on various graphs with structures like the hexagonal lattice, the square-octagon lattice, etc.~\cite{bombin2006topological, anderson2013homological}.
Here, for a positive integer $L$, we consider the $\llbracket 18 L^2, 4, 4L \rrbracket$ hexagonal color code $\cC$ with $\cC_X=\cC_Z$ based on a $3L\times 3L$ grid lying on a torus~\cite{sarvepalli2012efficient}.
For $L=2$, the relevant grid is shown in Fig.~\ref{fig:color_C_X_dual2}.
The drawing conventions are listed in Table~\ref{table:drawing:conventions:2}. 
W.l.o.g., we only consider the $X$ part of errors when decoding since $\cC_X=\cC_Z$.
\begin{table}[t]
\caption{Drawing conventions for figures of color codes.}
\label{table:drawing:conventions:2}
\begin{center}
\begin{tabular}{|l|l|}
\hline 
object & meaning\\
\hline \hline
empty vertex & $s_i = 0$ for syndrome bit for the $i^{\Th}$ parity check \\
\hline
filled vertex  & $s_i = 1$ for syndrome bit for the $i^{\Th}$ parity check \\
\hline
gray triangle & error introduced by the channel at that location \\
\hline
solid edge & part of output paths from a decoder for cycle codes\\
\hline
dashed edge & not part of output paths from a decoder for cycle codes\\
\hline
$e_{i, j}$ & edge whose endpoints are vertices labeled $i$ and $j$\\
\hline
$v_{i, j, k}$ & triangular face with vertices labeled $i$, $j$, and $k$\\
\hline
$c(i)$ & color of check/edge $i$ ($\Rm{R}$: red, $\Rm{G}$: green, and $\Rm{B}$: blue)\\
\hline
\end{tabular}
\end{center}
\end{table}
The color code $\cC$ is defined to be a CSS code based on ${\cG}$ in Fig.~\ref{fig:color_C_X_dual2}, where the codeword components are associated with the triangular faces and where checks are defined based on the incidence of the triangular faces on the vertices as specified further below in~\eqref{eq:def_H}.
The binary and quaternary stabilizer label codes of $\cC$ are, respectively, 
\begin{equation}
\tilde{\cB} \defeq \cC_X^{\perp} \times \cC_X^{\perp} \text{ and }\cB \defeq \GF{4}\text{-rowspan}(H),
\end{equation} 
where $H$ is the parity-check matrix of the $(3,6)$-regular LDPC code ${\cC}_X$ with  
\begin{equation}
\label{eq:def_H}
[H]_{i,j} \defeq [\text{ triangular face $j$ is incident on vertex $i$ }].
\end{equation}
Here $[S] \defeq 1$ when $S$ is true and $[S] \defeq 0$ otherwise for any statement $S$.
The code $\cC_X$ can be projected onto three cycle codes ${\cC}_{\Rm{R}}$, ${\cC}_{\Rm{G}}$, and ${\cC}_{\Rm{B}}$, 
where for $\Rm{C}\in \{\Rm{R}, \Rm{G}, \Rm{B}\}$, the parity-check matrix ${H}_{\Rm{C}}$ of ${\cC}_{\Rm{C}}$ is defined to be 
\begin{equation}
[{H}_{\Rm{C}}]_{i,j}\defeq [\text{ edge $j$ is incident on vertex $i$ }] \cdot [c(j)=\Rm{C}].
\end{equation}
Here, $\Rm{R}$, $\Rm{G}$, and $\Rm{B}$ stand for red, green, and blue, respectively.
For $\Rm{C}\in\{\Rm{R}, \Rm{G}, \Rm{B}\}$, let $\cG_{\Rm{C}}$ be the subgraph of $\cG$ in Fig.~\ref{fig:color_C_X_dual2} induced by the set of color-$\Rm{C}$ edges; replacing each vertex of $\cG_{\Rm{C}}$ by a single-parity check of degree three results in a normal factor graph for $\cC_{\Rm{C}}$.
For example, the graph ${\cG}_{\Rm{R}}$ for ${\cC}_{\Rm{R}}$ is shown in Fig.~\ref{fig:color_C_X_dual_R}.
For $\Rm{C}\in\{\Rm{R}, \Rm{G}, \Rm{B}\}$, let $f_{\Rm{C}}$ be the linear mapping described by the matrix $M_{\Rm{C}}$ such that
\begin{align}
&f_{\Rm{C}}:\ {\cC_X} \rightarrow \cC_{\Rm{C}},\ \vv \mapsto  \vv M_{\Rm{C}}^{\tr}, \\
&[M_{\Rm{C}}]_{i,j} \defeq [\text{edge $i$ is a color-$\Rm{C}$  edge of triangular face $j$}].
\end{align}
The combined vector of $f_{\Rm{C}}(\vv)$, $\Rm{C}\in\{\Rm{R}, \Rm{G}, \Rm{B}\}$, in $\cG$ is called the lifting of $\vv$ and is visualized for interesting scenarios in the upcoming Figs.~\ref{fig:color_eg1}-\ref{fig:color_eg1_v}. 
Suppose $\vs$ is the syndrome corresponding to the vector $\vv$.
One can check that 
\begin{equation}
\vs^{\tr} = H\vv^{\tr} \iff \vs^{\tr}_{\Rm{C}} =  {H}_{\Rm{C}} f_{\Rm{C}}(\vv)^{\tr}\ \forall \Rm{C},
\end{equation}
where $\vs_{\Rm{C}}$ is the corresponding syndrome in the induced subgraph $\cG_{\Rm{C}}$ of $\cG$ for $\Rm{C}\in\{\Rm{R}, \Rm{G}, \Rm{B}\}$.

\section{Pseudocodeword-based Decoding}
\label{section3:PCWDdecoding}

Decoding color codes by projecting $\cC_X$ onto cycle codes was first proposed in~\cite{delfosse2014decoding} and was improved in~\cite{kubica2019efficient}. 
The decoding procedure given the syndrome $\vs$ for $\cC$ in~\cite{delfosse2014decoding} is sketched as follows.
\begin{enumerate}
\item Compute $\vs_{\Rm{C}}$,
where $\vs_{\Rm{C}}$ is the subvector of $\vs$ induced by the subgraph $\cG_{\Rm{C}}$ of $\cG$ for $\Rm{C}\in\{\Rm{R}, \Rm{G}, \Rm{B}\}$.
\item For $\Rm{C}\in\{\Rm{R}, \Rm{G}, \Rm{B}\}$, decode w.r.t. ${\cC}_{\Rm{C}}$ to find a vector $\vv_{\Rm{C}}$ s.t.
\begin{equation}
\vs^{\tr}_{\Rm{C}} =  {H}_{\Rm{C}} {\vv}_{\Rm{C}}^{\tr}.
\end{equation}
\item Obtain the combined vector $\vv_{\Rm{RGB}}$ of $\vv_{\Rm{C}}$, $\Rm{C}\in\{\Rm{R}, \Rm{G}, \Rm{B}\}$, in $\cG$;
\item Find $\vv\in\cC_{X}$ s.t. the lifting of $\vv$ is $\vv_{\Rm{RGB}}$ 
and declare failure NOLIFTING if $\vv$ does not exist.
\end{enumerate}
Note that the fourth step in~\cite{kubica2019efficient} is simplified to  finding the vector $\vv$ s.t. $f_{\Rm{C}}(\vv) = \vv_{\Rm{C}}$ for $\Rm{C}\in\{\Rm{G}, \Rm{B}\}$.
\begin{figure}[t]
\begin{center}
\begin{tikzpicture}[x=1cm, y= 1cm, node/.style={draw=none},
factor/.style={draw,circle,fill=white, minimum size = 9pt},
factorR/.style={draw,circle, fill=none, minimum size = 9pt,color=red},
factorB/.style={draw,circle, fill=none, minimum size = 9pt,color=blue},
factorG/.style={draw,circle, fill=none, minimum size = 9pt,color=green},
factorRf/.style={draw,circle, fill=red, minimum size = 9pt,color=red},
factorBf/.style={draw,circle, fill=blue, minimum size = 9pt,color=blue},
factorGf/.style={draw,circle, fill=green, minimum size = 9pt,color=green}
]

\fill[fill=gray!20] (2,-1) -- (3,-2) -- (4,-2) -- (3,-1);
\fill[fill=gray!20] (4,-1) -- (4,-2) -- (5,-2);
\fill[fill=gray!20] (5,-1) -- (6,-1) -- (6,-2);
\node[factor] at (3,-1) {};
\node[factor] at (3,-2) {};
\node[factor] at (4,-2) {};
                          
\foreach \i in {0,1,2,3,4,5,6}{
\foreach \j in {0,1,2,3}{ 
\pgfmathparse{int(mod(\i+\j,3))} \let\tf\pgfmathresult 
\ifthenelse{\tf=0}{
	\node[factorR] (node\i\j) at (\i,-\j) {};
}{}; 
\ifthenelse{\tf=1}{
	\node[factorB] (node\i\j) at (\i,-\j) {};
}{}; 
\ifthenelse{\tf=2}{
	\node[factorG] (node\i\j) at (\i,-\j) {};
}{};  
}}                          
       
\foreach \i in {0,1,2,3,4,5}{
\foreach \j in {0,1,2}{
\pgfmathparse{int(mod(\i+\j,3))} \let\tf\pgfmathresult 
\pgfmathtruncatemacro\jj{\j+1}; 
\pgfmathtruncatemacro\ii{\i+1}; 
\ifthenelse{\tf=0}{
		\draw[green,dashed] (node\i\j) -- (node\i\jj); 
		\draw[green,dashed] (node\i\j) -- (node\ii\j);
		\draw[blue,dashed] (node\i\j) -- (node\ii\jj); 
}{}
\ifthenelse{\tf=1}{
		\draw[red,dashed] (node\i\j) -- (node\i\jj); 
		\draw[red,dashed] (node\i\j) -- (node\ii\j);
		\draw[green,dashed] (node\i\j) -- (node\ii\jj); 
}{}
\ifthenelse{\tf=2}{
		\draw[blue,dashed] (node\i\j) -- (node\i\jj); 
		\draw[blue,dashed] (node\i\j) -- (node\ii\j); 
		\draw[red,dashed] (node\i\j) -- (node\ii\jj); 
}{}
}}     
              
\foreach \i in {6}{
\foreach \j in {0,1,2}{   
\pgfmathparse{int(mod(\i+\j,3))} \let\tf\pgfmathresult 
\pgfmathtruncatemacro\jj{\j+1}; 
\ifthenelse{\tf=0}{
		\draw[green,dashed] (node\i\j) -- (node\i\jj); 
}{}
\ifthenelse{\tf=1}{
		\draw[red,dashed] (node\i\j) -- (node\i\jj); 
}{}
\ifthenelse{\tf=2}{
		\draw[blue,dashed] (node\i\j) -- (node\i\jj); 
}{}
}}    

\foreach \i in {0,1,2,3,4,5}{
\foreach \j in {3}{
\pgfmathparse{int(mod(\i+\j,3))} \let\tf\pgfmathresult 
\pgfmathtruncatemacro\ii{\i+1}; 
\ifthenelse{\tf=0}{
		\draw[green,dashed] (node\i\j) -- (node\ii\j);
}{}
\ifthenelse{\tf=1}{
		\draw[red,dashed] (node\i\j) -- (node\ii\j);
}{}
\ifthenelse{\tf=2}{
		\draw[blue,dashed] (node\i\j) -- (node\ii\j);
}{}
}}

\node[factorRf] at (2,-1) {};
\node[factorRf] at (5,-1) {};
\node[factorGf] at (4,-1) {};
\node[factorGf] at (6,-2) {};
\node[factorGf] at (0,-2) {};
\node[factorBf] at (5,-2) {};
\node[factorBf] at (6,-1) {};
\node[factorBf] at (0,-1) {};

\draw[blue,thick] (node02) -- (node12);
\draw[blue,thick] (node12) -- (node11);
\draw[blue,thick] (node11) -- (node21);
\draw[green,thick] (node21) -- (node31);
\draw[green,thick] (node31) -- (node42);
\draw[green,thick] (node42) -- (node52);
\draw[red,thick] (node52) -- (node41);
\draw[blue,thick] (node41) -- (node51);
\draw[green,thick] (node51) -- (node61);
\draw[red,thick] (node61) -- (node62);
\end{tikzpicture}
\end{center}
\caption{An example of output paths (solid edges) of the decoder for cycle codes leading to decoding failure for the decoder in~\cite{delfosse2014decoding}, where only the relevant part is shown.}
  \label{fig:color_eg1}
\end{figure}
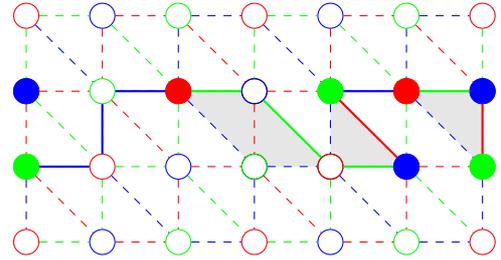
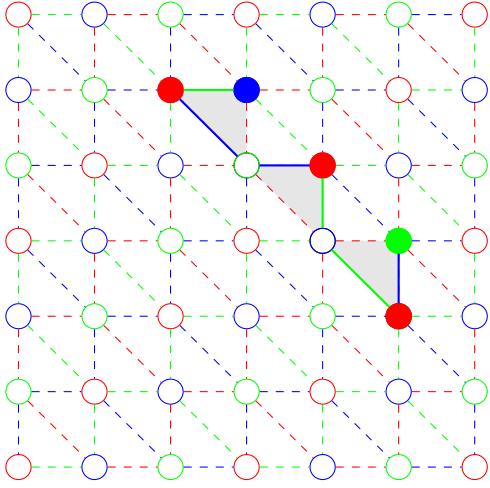
\begin{figure}[t]
\begin{center}
\begin{tikzpicture}[x=1cm, y= 1cm, node/.style={draw=none},
factor/.style={draw,circle,fill=white, minimum size = 9pt},
factorR/.style={draw,circle, fill=none, minimum size = 9pt,color=red},
factorB/.style={draw,circle, fill=none, minimum size = 9pt,color=blue},
factorG/.style={draw,circle, fill=none, minimum size = 9pt,color=green},
factorRf/.style={draw,circle, fill=red, minimum size = 9pt,color=red},
factorBf/.style={draw,circle, fill=blue, minimum size = 9pt,color=blue},
factorGf/.style={draw,circle, fill=green, minimum size = 9pt,color=green}
]

\fill[fill=gray!20] (2,-1) -- (3,-1) -- (3,-2);
\fill[fill=gray!20] (3,-2) -- (4,-2) -- (4,-3);
\fill[fill=gray!20] (4,-3) -- (5,-3) -- (5,-4);
\node[factor] at (3,-2) {};
\node[factor] at (4,-3) {};
\node[factor] at (5,-4) {};
                          
\foreach \i in {0,1,2,3,4,5,6}{ 
\foreach \j in {0,1,2,3,4,5,6}{ 
\pgfmathparse{int(mod(\i+\j,3))} \let\tf\pgfmathresult 
\ifthenelse{\tf=0}{
	\node[factorR] (node\i\j) at (\i,-\j) {};
}{}; 
\ifthenelse{\tf=1}{
	\node[factorB] (node\i\j) at (\i,-\j) {};
}{}; 
\ifthenelse{\tf=2}{
	\node[factorG] (node\i\j) at (\i,-\j) {};
}{};  
}}                          
       
\foreach \i in {0,1,2,3,4,5}{
\foreach \j in {0,1,2,3,4,5}{
\pgfmathparse{int(mod(\i+\j,3))} \let\tf\pgfmathresult 
\pgfmathtruncatemacro\jj{\j+1}; 
\pgfmathtruncatemacro\ii{\i+1}; 
\ifthenelse{\tf=0}{
		\draw[green,dashed] (node\i\j) -- (node\i\jj); 
		\draw[green,dashed] (node\i\j) -- (node\ii\j);
		\draw[blue,dashed] (node\i\j) -- (node\ii\jj); 
}{}
\ifthenelse{\tf=1}{
		\draw[red,dashed] (node\i\j) -- (node\i\jj); 
		\draw[red,dashed] (node\i\j) -- (node\ii\j);
		\draw[green,dashed] (node\i\j) -- (node\ii\jj); 
}{}
\ifthenelse{\tf=2}{
		\draw[blue,dashed] (node\i\j) -- (node\i\jj); 
		\draw[blue,dashed] (node\i\j) -- (node\ii\j); 
		\draw[red,dashed] (node\i\j) -- (node\ii\jj); 
}{}
}}     
              
\foreach \i in {6}{
\foreach \j in {0,1,2,3,4,5}{   
\pgfmathparse{int(mod(\i+\j,3))} \let\tf\pgfmathresult 
\pgfmathtruncatemacro\jj{\j+1}; 
\ifthenelse{\tf=0}{
		\draw[green,dashed] (node\i\j) -- (node\i\jj); 
}{}
\ifthenelse{\tf=1}{
		\draw[red,dashed] (node\i\j) -- (node\i\jj); 
}{}
\ifthenelse{\tf=2}{
		\draw[blue,dashed] (node\i\j) -- (node\i\jj); 
}{}
}}    

\foreach \i in {0,1,2,3,4,5}{
\foreach \j in {6}{
\pgfmathparse{int(mod(\i+\j,3))} \let\tf\pgfmathresult 
\pgfmathtruncatemacro\ii{\i+1}; 
\ifthenelse{\tf=0}{
		\draw[green,dashed] (node\i\j) -- (node\ii\j);
}{}
\ifthenelse{\tf=1}{
		\draw[red,dashed] (node\i\j) -- (node\ii\j);
}{}
\ifthenelse{\tf=2}{
		\draw[blue,dashed] (node\i\j) -- (node\ii\j);
}{}
}}         

\node[factorRf] at (2,-1) {};
\node[factorBf] at (3,-1) {};
\node[factorRf] at (4,-2) {};
\node[factorGf] at (5,-3) {};
\node[factorRf] at (5,-4) {};

\draw[green,thick] (node21) -- (node31);

\draw[blue,thick] (node53) -- (node54);

\draw[blue,thick] (node21) -- (node32);
\draw[blue,thick] (node32) -- (node42);

\draw[green,thick] (node42) -- (node43);
\draw[green,thick] (node43) -- (node54);
\end{tikzpicture}
\end{center}
\caption{An example of output paths (solid edges) of the decoder for cycle codes leading to decoding failure for the decoder in~\cite{kubica2019efficient}.}
  \label{fig:color_eg2}
\end{figure}
The issues of the previous approaches are as follows.
\begin{itemize}
\item Finding the vector given its lifting is a global operation of high computational complexity in~\cite{delfosse2014decoding}, which is improved to be a local operation of low computational complexity in~\cite{kubica2019efficient}. 
\item Both approaches in~\cite{delfosse2014decoding,kubica2019efficient} use the output of decoders for cycle codes, e.g., the MWPM algorithm, which may lead to decoding failure NOLIFTING since some paths are dropped by the decoders for cycle codes.
\begin{itemize}
\item For example, the solid edges in Fig.~\ref{fig:color_eg1} are possible output paths from a decoder for cycle codes. 
However, the decoder in~\cite{delfosse2014decoding} will not find a vector whose lifting is associated with the solid edges since the solid edges cannot form a boundary of some region consisting of triangular faces.
\item For example, the solid edges in Fig.~\ref{fig:color_eg2} are possible output paths from a decoder for cycle codes ${\cC}_{\Rm{B}}$ and ${\cC}_{\Rm{G}}$, which will lead to decoding failure for the decoder in~\cite{kubica2019efficient}, 
since the two paths with endpoints of different colors cannot be converted into vectors because of no shared endpoint.
\end{itemize}
\end{itemize}

Based on the idea of the two-stage pseudocodeword-based decoder for quantum cycle codes, we introduce a two-stage  pseudocodeword-based decoder for color codes, named SPA+LPPCWD for color codes in Algorithm~\ref{alg:SPA_LPPCWD_color} with ``generalized paths'', i.e., vectors in $\cC_X$. 
First stage: run the SPA for $\cC_X$. 
Second stage: if the SPA succeeds, then stop; otherwise, run the SPA on $\cG_{\Rm{R}}$, $\cG_{\Rm{G}}$, and $\cG_{\Rm{B}}$, decompose the SPA pseudocodewords into paths, convert paths into ``generalized paths'', and finally use an (integer) LP to choose some collection of ``generalized paths'' to obtain a vector with a matching syndrome. 
Note that we only use part of the pseudocodeword-based decoder for cycle codes, i.e., the PCWD for cycle codes in Algorithm~\ref{alg:PCWD}, to find paths without selection.
The main challenge is to generalize the paths obtained from Algorithm~\ref{alg:PCWD} to obtain vectors in color codes $\cC_X$ ``generalized paths''.
We use a method to convert paths into vectors, which is similar to that in~\cite{kubica2019efficient} and is also a local operation of low computational complexity, but the conversion operation is done w.r.t. the three cycle codes $\cC_{\Rm{R}}$, $\cC_{\Rm{G}}$, and $\cC_{\Rm{B}}$ separately instead of only considering two cycle codes $\cC_{\Rm{G}}$ and $\cC_{\Rm{B}}$ as in~\cite{kubica2019efficient}.
Note that these changes compared with previous decoding approaches help to avoid problems as in Fig.~\ref{fig:color_eg2}.
Finally, we use an LP to choose a suitable collection of ``generalized paths'' to obtain a vector with a matching syndrome, where an integer LP can be used when LP fails to output an integer point, which guarantees that we can always obtain a vector with a matching syndrome of the minimum Hamming weight among all such vectors and also avoids problems as in Fig.~\ref{fig:color_eg1} and Fig.~\ref{fig:color_eg2}.
\begin{algorithm}[t]
\caption{Two-stage pseudocodeword-based decoder (SPA+LPPCWD) for color codes} 
\label{alg:SPA_LPPCWD_color} 
\begin{algorithmic}[1]
\REQUIRE the syndrome $\vs$ and the max.\ number of SPA iterations.
\ENSURE $\vv+\cB$. 
\STATE Run the SPA for the color code and obtain a vector $\vv$.
\IF {$H \vv^{\tr} = \vs^{\tr}$}
	\STATE Return $\vv+\cB$. 
	\COMMENT{SPA for the color code succeeds.}
\ELSE
\STATE Obtain the subvector $\vs_{\Rm{C}}$ of $\vs$ induced by the subgraph $\cG_{\Rm{C}}$ of $\cG$ for $\Rm{C}\in\{\Rm{R}, \Rm{G}, \Rm{B}\}$.
\STATE Run the SPA on NFGs for $\cC_{\Rm{R}}$, $\cC_{\Rm{B}}$, and $\cC_{\Rm{G}}$ separately to obtain three SPA pseudocodewords $\boldsymbol{\om}_{\Rm{C}}$, $\Rm{C}\in\{\Rm{R}, \Rm{G}, \Rm{B}\}$.
\STATE Initiate $P\leftarrow \emptyset$, $S\leftarrow \emptyset$, $V\leftarrow \emptyset$, and $S'\leftarrow \emptyset$.
\COMMENT{$P$ is the set of paths, $S$ and $S'$ are the sets of unsatisfied checks, and $V$ is the set of supports of vectors.}
\STATE Obtain $P \leftarrow P \cup \{P_j\}$, $S \leftarrow S \cup \{S_j\}$ by Algorithm~\ref{alg:PCWD} with input $\boldsymbol{\om}_{\Rm{C}}$ for $\Rm{C}\in\{\Rm{R}, \Rm{G}, \Rm{B}\}$.
\FOR{
$i$ with $|c(S_i)|=1$ or $i \neq j$ with $|c(S_i)|=2$, $|c(S_i \cup S_j)| = 3$, and $|S_i \cup S_j|=3$
}
\STATE Convert $P_i$ (together with $P_j$) into a vector $\vv_{\ell}$ with the set of unsatisfied checks $S'_{\ell}$, $V_{\ell} \defeq \Supp(\vv_{\ell})$, and cost $\lambda_{\ell} \defeq |V_{\ell}|$. $V \leftarrow V \cup \{V_{\ell}\}$ and $S' \leftarrow S' \cup \{S'_{\ell}\}$.
\ENDFOR
\STATE Use an (integer) LP to choose a collection of vectors from $V$ to obtain a vector with a matching syndrome. 
\ENDIF
\end{algorithmic}
\end{algorithm}
\begin{algorithm}[t]
\caption{Pseudocodeword decomposition (PCWD) for cycle codes} 
\label{alg:PCWD} 
\begin{algorithmic}[1]
\REQUIRE a pseudocodeword $\boldsymbol{\om}$ and the syndrome $\vs$.
\ENSURE $P = \{P_i\}$, $S=\{S_i\}$, and a weight vector $\hat{\boldV{\om}}$.
\STATE $\cJ \leftarrow  \{j\ |\ s_j\!\neq\! 0\}$ 
\COMMENT{The set of unsatisfied checks.}
\STATE $P \leftarrow \emptyset$ 
\COMMENT{A set of paths.}
\STATE $S \leftarrow \emptyset$
\COMMENT{A set of unsatisfied checks for each path.}
\WHILE{$\cJ \neq \emptyset$}
	\STATE For each $j\in \cJ$, set $\overline{\om}_j \leftarrow 0$ when $s_j$ is isolated. Otherwise, start from $s_j$ and follow the edge with the largest possible component of $\boldsymbol{\om}$ at each step without repetition until reaching $s_{j'}$, $j' \in \cJ$, to obtain a path/cycle $P_j$ with weight $\overline{\om}_j \leftarrow \min_{\ell \in P_j} \om_\ell$ and $S_j \leftarrow \{j,j'\}$. 
	\STATE $i \leftarrow \argmin_{j} (1-\overline{\om}_j) \cdot |P_j|$. 
	\COMMENT{The min.-cost path.}
	\STATE $\om_\ell \leftarrow \om_\ell - \overline{\om}_{i}$ $\forall \ell \in P_{i}$. 
	\COMMENT{Remove $P_i$ from $\boldsymbol{\om}$.}
\IF{$|S_i|>1$}
	\STATE $P \leftarrow P \cup \{ P_{i}\}$, $S \leftarrow S \cup \{ S_{i}\}$, $\ell \leftarrow |P|$, and $\hat{\om}_{\ell} \leftarrow \overline{\om}_i$.
	\COMMENT{Include path $P_i$ in $P$.}
\ENDIF
	
	\STATE $\cJ \leftarrow \cJ \backslash \{j \in \cJ\ |\ \overline{\om}_j=0\}$.
	\COMMENT{Remove isolated checks.}
\ENDWHILE
\end{algorithmic}
\end{algorithm}

\begin{figure}[t]
\begin{center}
\resizebox{0.7\columnwidth}{!}{\begin{tikzpicture}[x=1cm, y= 1cm, node/.style={draw=none},
factor/.style={draw,circle,fill=white, minimum size = 11pt},
factorR/.style={draw,circle, fill=none, minimum size = 11pt,color=red},
factorB/.style={draw,circle, fill=none, minimum size = 11pt,color=blue},
factorG/.style={draw,circle, fill=none, minimum size = 11pt,color=green},
factorRf/.style={draw,circle, fill=red, minimum size = 11pt,color=red},
factorBf/.style={draw,circle, fill=blue, minimum size = 11pt,color=blue},
factorGf/.style={draw,circle, fill=green, minimum size = 11pt,color=green}
]

\foreach \i in {0,1,2,3,4,5,6}{ 
	\foreach \j in {0,1,2,3,4,5,6}{ 
		\node (node\i\j) at (\i,-\j) {};
}}  

\fill[fill=gray!20] (3,-1) -- (3,-2) -- (4,-2);
\fill[fill=gray!20] (3,-2) -- (3,-3) -- (4,-3);
\fill[fill=gray!20] (4,-2) -- (4,-3) -- (5,-3);
\fill[fill=gray!20] (1,-4) -- (2,-4) -- (2,-5);
    
\foreach \i in {0,1,2,3,4,5}{
\foreach \j in {0,1,2,3,4,5}{
\pgfmathparse{int(mod(\i+\j,3))} \let\tf\pgfmathresult 
\pgfmathtruncatemacro\jj{\j+1}; 
\pgfmathtruncatemacro\ii{\i+1}; 
\ifthenelse{\tf=0}{
		\draw[green,dashed] (node\i\j) -- (node\i\jj); 
		\draw[green,dashed] (node\i\j) -- (node\ii\j);
		\draw[blue,dashed] (node\i\j) -- (node\ii\jj); 
}{}
\ifthenelse{\tf=1}{
		\draw[red,dashed] (node\i\j) -- (node\i\jj); 
		\draw[red,dashed] (node\i\j) -- (node\ii\j);
		\draw[green,dashed] (node\i\j) -- (node\ii\jj); 
}{}
\ifthenelse{\tf=2}{
		\draw[blue,dashed] (node\i\j) -- (node\i\jj); 
		\draw[blue,dashed] (node\i\j) -- (node\ii\j); 
		\draw[red,dashed] (node\i\j) -- (node\ii\jj); 
}{}
}}     
              
\foreach \i in {6}{
\foreach \j in {0,1,2,3,4,5}{   
\pgfmathparse{int(mod(\i+\j,3))} \let\tf\pgfmathresult 
\pgfmathtruncatemacro\jj{\j+1}; 
\ifthenelse{\tf=0}{
		\draw[green,dashed] (node\i\j) -- (node\i\jj); 
}{}
\ifthenelse{\tf=1}{
		\draw[red,dashed] (node\i\j) -- (node\i\jj); 
}{}
\ifthenelse{\tf=2}{
		\draw[blue,dashed] (node\i\j) -- (node\i\jj); 
}{}
}}    

\foreach \i in {0,1,2,3,4,5}{
\foreach \j in {6}{
\pgfmathparse{int(mod(\i+\j,3))} \let\tf\pgfmathresult 
\pgfmathtruncatemacro\ii{\i+1}; 
\ifthenelse{\tf=0}{
		\draw[green,dashed] (node\i\j) -- (node\ii\j);
}{}
\ifthenelse{\tf=1}{
		\draw[red,dashed] (node\i\j) -- (node\ii\j);
}{}
\ifthenelse{\tf=2}{
		\draw[blue,dashed] (node\i\j) -- (node\ii\j);
}{}
}}  

\draw[blue,thick] (node14) -- (node24);
\draw[green,thick] (node24) -- (node25);
\draw[red,thick] (node14) -- (node25);
\draw[red,thick] (node31) -- (node32) -- (node43) -- (node53);
\draw[blue,thick] (node33) -- (node32) -- (node42) -- (node53);
\draw[green,thick] (node31) -- (node42) -- (node43) -- (node33);
\draw[green,thick] (node30) -- (node31);
\draw[green,thick] (node25) -- (node36);
\draw[blue,thick] (node33) -- (node23) -- (node24);
\draw[green,thick] (node33) -- (node34) -- (node24);
\draw[red,thick] (node53) -- (node64);
\draw[red,thick] (node04) -- (node14);
\draw[blue,thick] (node53) -- (node63);
\draw[blue,thick] (node03) -- (node14);
\draw[red,thick] (node25) -- (node26);
\draw[red,thick] (node20) -- (node31);

\foreach \i in {0,1,2,3,4,5,6}{ 
	\foreach \j in {0,1,2,3,4,5,6}{ 
		\node[factor] (node\i\j) at (\i,-\j) {};
}}  

\foreach \i in {0,1,2,3,4,5,6}{ 
\foreach \j in {0,1,2,3,4,5,6}{ 
\pgfmathparse{int(mod(\i+\j,3))} \let\tf\pgfmathresult 
\ifthenelse{\tf=0}{
	\node[factorR] (node\i\j) at (\i,-\j) {};
}{}; 
\ifthenelse{\tf=1}{
	\node[factorB] (node\i\j) at (\i,-\j) {};
}{}; 
\ifthenelse{\tf=2}{
	\node[factorG] (node\i\j) at (\i,-\j) {};
}{};  
}}               

\node at (2,0) {\small$1$};
\node at (2,-6) {\small$1$};
\node at (3,0) {\small$2$};
\node at (3,-6) {\small$2$};
\node[factorBf] at (3,-1) {};
\node at (3,-1) {\small\color{white}$3$};
\node at (3,-2) {\small$4$};
\node at (4,-2) {\small$5$};
\node at (0,-3) {\small$6$};
\node at (6,-3) {\small$6$};
\node at (2,-3) {\small$7$};
\node[factorRf] at (3,-3) {};
\node at (3,-3) {\small\color{white}$8$};
\node at (4,-3) {\small$9$};
\node[factorGf] at (5,-3) {};
\node at (5,-3) {\small\color{white}$10$};
\node at (0,-4) {\small$11$};
\node at (6,-4) {\small$11$};
\node[factorGf] at (1,-4) {};
\node at (1,-4) {\small\color{white}$12$};
\node[factorRf] at (2,-4) {};
\node at (2,-4) {\small\color{white}$13$};
\node at (3,-4) {\small$14$};
\node[factorBf] at (2,-5) {};
\node at (2,-5) {\small\color{white}$15$};

\end{tikzpicture}}
\end{center}
\caption{Output paths of Algorithm~\ref{alg:PCWD} in Example~\ref{eg:SPA_LPPCWD_coc_eg}.}
\label{fig:color_eg1_path}
\end{figure}
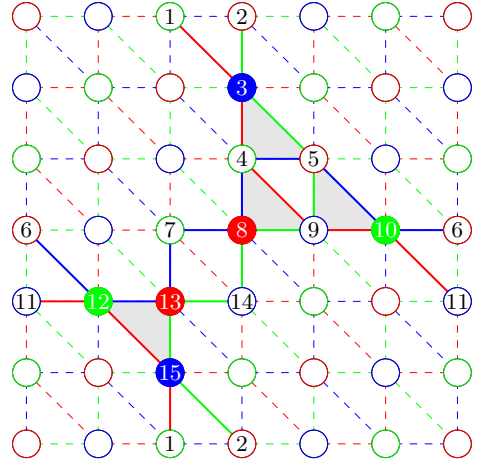
\begin{figure}[t]
\begin{center}
\resizebox{0.7\columnwidth}{!}{\begin{tikzpicture}[x=1cm, y= 1cm, node/.style={draw=none},
factor/.style={draw,circle,fill=white, minimum size = 11pt},
factorR/.style={draw,circle, fill=none, minimum size = 11pt,color=red},
factorB/.style={draw,circle, fill=none, minimum size = 11pt,color=blue},
factorG/.style={draw,circle, fill=none, minimum size = 11pt,color=green},
factorRf/.style={draw,circle, fill=red, minimum size = 11pt,color=red},
factorBf/.style={draw,circle, fill=blue, minimum size = 11pt,color=blue},
factorGf/.style={draw,circle, fill=green, minimum size = 11pt,color=green}
]
\definecolor{BlueViolet}{RGB}{138,43,226}
\definecolor{tomato}{RGB}{255,99,71}
\definecolor{LimeGreen}{RGB}{50,205,50}

\foreach \i in {0,1,2,3,4,5,6}{ 
	\foreach \j in {0,1,2,3,4,5,6}{ 
		\node (node\i\j) at (\i,-\j) {};
}}  

\path[draw, pattern=horizontal lines, pattern color=BlueViolet] (3,-1) -- (3,-2) -- (4,-2);
\path[draw, pattern=horizontal lines, pattern color=BlueViolet] (3,-2) -- (3,-3) -- (4,-3);
\path[draw, pattern=horizontal lines, pattern color=BlueViolet] (4,-2) -- (4,-3) -- (5,-3);
\path[draw, pattern=vertical lines, pattern color=blue] (1,-4) -- (2,-4) -- (2,-5);
\path[draw, pattern=grid, pattern color=tomato] (2,0) -- (3,0) -- (3,-1); 
\path[draw, pattern=grid, pattern color=tomato] (2,-5) -- (2,-6) -- (3,-6);
\path[draw, pattern=crosshatch, pattern color=LimeGreen] (2,-3) -- (3,-3) -- (3,-4) -- (2,-4); 
\path[draw, pattern=fivepointed stars, pattern color=orange] (0,-3) -- (0,-4) -- (1,-4); 
\path[draw, pattern=fivepointed stars, pattern color=orange] (5,-3) -- (6,-3) -- (6,-4);
    
\foreach \i in {0,1,2,3,4,5}{
\foreach \j in {0,1,2,3,4,5}{
\pgfmathparse{int(mod(\i+\j,3))} \let\tf\pgfmathresult 
\pgfmathtruncatemacro\jj{\j+1}; 
\pgfmathtruncatemacro\ii{\i+1}; 
\ifthenelse{\tf=0}{
		\draw[green,dashed] (node\i\j) -- (node\i\jj); 
		\draw[green,dashed] (node\i\j) -- (node\ii\j);
		\draw[blue,dashed] (node\i\j) -- (node\ii\jj); 
}{}
\ifthenelse{\tf=1}{
		\draw[red,dashed] (node\i\j) -- (node\i\jj); 
		\draw[red,dashed] (node\i\j) -- (node\ii\j);
		\draw[green,dashed] (node\i\j) -- (node\ii\jj); 
}{}
\ifthenelse{\tf=2}{
		\draw[blue,dashed] (node\i\j) -- (node\i\jj); 
		\draw[blue,dashed] (node\i\j) -- (node\ii\j); 
		\draw[red,dashed] (node\i\j) -- (node\ii\jj); 
}{}
}}     
              
\foreach \i in {6}{
\foreach \j in {0,1,2,3,4,5}{   
\pgfmathparse{int(mod(\i+\j,3))} \let\tf\pgfmathresult 
\pgfmathtruncatemacro\jj{\j+1}; 
\ifthenelse{\tf=0}{
		\draw[green,dashed] (node\i\j) -- (node\i\jj); 
}{}
\ifthenelse{\tf=1}{
		\draw[red,dashed] (node\i\j) -- (node\i\jj); 
}{}
\ifthenelse{\tf=2}{
		\draw[blue,dashed] (node\i\j) -- (node\i\jj); 
}{}
}}    

\foreach \i in {0,1,2,3,4,5}{
\foreach \j in {6}{
\pgfmathparse{int(mod(\i+\j,3))} \let\tf\pgfmathresult 
\pgfmathtruncatemacro\ii{\i+1}; 
\ifthenelse{\tf=0}{
		\draw[green,dashed] (node\i\j) -- (node\ii\j);
}{}
\ifthenelse{\tf=1}{
		\draw[red,dashed] (node\i\j) -- (node\ii\j);
}{}
\ifthenelse{\tf=2}{
		\draw[blue,dashed] (node\i\j) -- (node\ii\j);
}{}
}}  

\draw[blue,thick] (node14) -- (node24);
\draw[green,thick] (node24) -- (node25);
\draw[red,thick] (node14) -- (node25);
\draw[red,thick] (node31) -- (node32) -- (node43) -- (node53);
\draw[blue,thick] (node33) -- (node32) -- (node42) -- (node53);
\draw[green,thick] (node31) -- (node42) -- (node43) -- (node33);
\draw[green,thick] (node30) -- (node31);
\draw[green,thick] (node25) -- (node36);
\draw[blue,thick] (node33) -- (node23) -- (node24);
\draw[green,thick] (node33) -- (node34) -- (node24);
\draw[red,thick] (node53) -- (node64);
\draw[red,thick] (node04) -- (node14);
\draw[blue,thick] (node53) -- (node63);
\draw[blue,thick] (node03) -- (node14);
\draw[red,thick] (node25) -- (node26);
\draw[red,thick] (node20) -- (node31);

\foreach \i in {0,1,2,3,4,5,6}{ 
	\foreach \j in {0,1,2,3,4,5,6}{ 
		\node[factor] (node\i\j) at (\i,-\j) {};
}}  

\foreach \i in {0,1,2,3,4,5,6}{ 
\foreach \j in {0,1,2,3,4,5,6}{ 
\pgfmathparse{int(mod(\i+\j,3))} \let\tf\pgfmathresult 
\ifthenelse{\tf=0}{
	\node[factorR] (node\i\j) at (\i,-\j) {};
}{}; 
\ifthenelse{\tf=1}{
	\node[factorB] (node\i\j) at (\i,-\j) {};
}{}; 
\ifthenelse{\tf=2}{
	\node[factorG] (node\i\j) at (\i,-\j) {};
}{};  
}}               

\node at (2,0) {\small$1$};
\node at (2,-6) {\small$1$};
\node at (3,0) {\small$2$};
\node at (3,-6) {\small$2$};
\node[factorBf] at (3,-1) {};
\node at (3,-1) {\small\color{white}$3$};
\node at (3,-2) {\small$4$};
\node at (4,-2) {\small$5$};
\node at (0,-3) {\small$6$};
\node at (6,-3) {\small$6$};
\node at (2,-3) {\small$7$};
\node[factorRf] at (3,-3) {};
\node at (3,-3) {\small\color{white}$8$};
\node at (4,-3) {\small$9$};
\node[factorGf] at (5,-3) {};
\node at (5,-3) {\small\color{white}$10$};
\node at (0,-4) {\small$11$};
\node at (6,-4) {\small$11$};
\node[factorGf] at (1,-4) {};
\node at (1,-4) {\small\color{white}$12$};
\node[factorRf] at (2,-4) {};
\node at (2,-4) {\small\color{white}$13$};
\node at (3,-4) {\small$14$};
\node[factorBf] at (2,-5) {};
\node at (2,-5) {\small\color{white}$15$};

\end{tikzpicture}}
\end{center}
\caption{Vectors converted from paths in Fig.~\ref{fig:color_eg1_path}.}
\label{fig:color_eg1_v}
\end{figure}
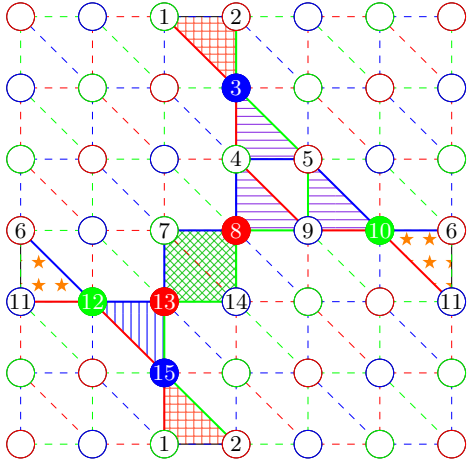
\begin{table}
\caption{Output of Algorithm~\ref{alg:PCWD} in Example~\ref{eg:SPA_LPPCWD_coc_eg}, 
where $c(P_i)$ and $c(S_i)$ are the colors of $P_i$ and its endpoints, respectively.}
\label{table:SPA_LPPCWD_coc_eg_PCWD}
\begin{center}
\begin{tabular}{| c | c | c | c | c | c |}
\hline
$i$ & $P_i$ & $\hat{\om}_i$ & $S_i$ & $c(P_i)$ & $c(S_i)$\\
\hline \hline
$1$ & $\{e_{12,15}\}$ & $0.7405$ &  $\{12,\ 15\}$ & $\{\Rm{R}\}$ & $\{\Rm{G}, \Rm{B}\}$\\
\hline
$2$ & $\{e_{3,1},\ e_{1,15}\}$ & $0.2196$ &  $\{3,\ 15\}$ & $\{\Rm{R}\}$ & $\{\Rm{B}\}$\\
\hline
$3$ & $\{e_{10,11},\ e_{11,12}\}$ & $0.2196$ &  $\{10,\ 12\}$ & $\{\Rm{R}\}$ & $\{\Rm{G}\}$\\
\hline
$4$ & $\{e_{3,4},\ e_{4,9},\ e_{9,10}\}$ & $0.2002$ &  $\{3,\ 10\}$ & $\{\Rm{R}\}$ & $\{\Rm{G},\Rm{B}\}$\\
\hline
$5$ & $\{e_{12,13}\}$ & $0.7405$ &  $\{12,\ 13\}$ & $\{\Rm{B}\}$ & $\{\Rm{R},\Rm{G}\}$\\
\hline
$6$ & $\{e_{8,7},\ e_{7,13}\}$ & $0.2196$ &  $\{8,\ 13\}$ & $\{\Rm{B}\}$ & $\{\Rm{R}\}$\\
\hline
$7$ & $\{e_{10,6},\ e_{6,12}\}$ & $0.2196$ &  $\{10,\ 12\}$ & $\{\Rm{B}\}$ & $\{\Rm{G}\}$\\
\hline
$8$ & $\{e_{8,4},\ e_{4,5},\ e_{5,10}\}$ & $0.2002$ &  $\{8,\ 10\}$ & $\{\Rm{B}\}$ & $\{\Rm{R},\Rm{G}\}$\\
\hline
$9$ & $\{e_{13,15}\}$ & $0.7405$ &  $\{13,\ 15\}$ & $\{\Rm{G}\}$ & $\{\Rm{R},\Rm{B}\}$\\
\hline
$10$ & $\{e_{3,2},\ e_{2,15}\}$ & $0.2196$ &  $\{3,\ 15\}$ & $\{\Rm{G}\}$ & $\{\Rm{B}\}$\\
\hline
$11$ & $\{e_{8,14},\ e_{14,13}\}$ & $0.2196$ &  $\{8,\ 13\}$ & $\{\Rm{G}\}$ & $\{\Rm{R}\}$\\
\hline
$12$ & $\{e_{3,5},\ e_{5,9},\ e_{9,8}\}$ & $0.2002$ &  $\{3,\ 8\}$ & $\{\Rm{G}\}$ & $\{\Rm{R},\Rm{B}\}$\\
\hline
\end{tabular}
\end{center}
\end{table}
\begin{table}
\caption{``Generalized paths'', i.e., vectors in $\cC_X$, converted from paths in Table~\ref{table:SPA_LPPCWD_coc_eg_PCWD}.} 
\label{table:SPA_LPPCWD_coc_eg_PCWD_v}
\begin{center}
\begin{tabular}{| c | c | c | c | c |}
\hline
$i$ & $V_i$ & $\lambda_i$ & $S'_i$ & Pattern\\
\hline \hline
$1$ & $\{v_{1,2,3},\, v_{1,2,15}\}$ & $2$ &  $\{3,\, 15\}$ & grid \\
\hline
$2$ & $\{v_{6,11,12},\, v_{6,10,11}\}$ & $2$ &  $\{10,\, 12\}$ & fivepointed stars \\
\hline
$3$ & $\{v_{7,8,14},\, v_{7,13,14}\}$ & $2$ &  $\{8,\, 13\}$ & crosshatch \\
\hline
$4$ & $\{v_{12,13,15}\}$ & $1$ &  $\{12,\, 13,\, 15\}$ & vertical lines \\
\hline
$5$ & $\{v_{3,4,5},\, v_{4,8,9},\, v_{5,9,10}\}$ & $3$ &  $\{3,\, 8,\, 10\}$ & horizontal lines \\
\hline
\end{tabular}
\end{center}
\end{table}
We use Example~\ref{eg:SPA_LPPCWD_coc_eg} to demonstrate how to decode color codes using Algorithm~\ref{alg:SPA_LPPCWD_color}.
\begin{example}
\label{eg:SPA_LPPCWD_coc_eg}
Consider the $\llbracket 72,4,8\rrbracket$ color code and the quantum depolarizing channel with depolarizing probability $p=0.0251$.
W.l.o.g, we only consider the X part of the error.
Suppose the errors happened at the locations corresponding to the gray triangles, namely $v_{3,4,5}$, $v_{4,8,9}$, $v_{5,9,10}$, and $v_{12,13,15}$, in Fig.~\ref{fig:color_eg1_path}, where only relevant vertices are labeled.
The syndrome obtained after measurements is the binary vector $\vs$ with support $\Supp(\vs) = \{3,8,10,12,13,15\}$. 
First, the SPA for $\cC_X$ fails. 
Then, the syndrome $\vs_{\Rm{R}}$ for $\cC_{\Rm{R}}$ is obtained by removing all the red vertices in $\vs$, namely removing $s_i$, $i\in\{2,5,6,8,13\}$.
In a similar manner, we can obtain the syndromes $\vs_{\Rm{G}}$ and $\vs_{\Rm{B}}$ (for $\cC_{\Rm{G}}$ and $\cC_{\Rm{B}}$, respectively) from $\vs$.
After running the SPA on NFGs for $\cC_{\Rm{R}}$, $\cC_{\Rm{G}}$, and $\cC_{\Rm{B}}$ separately, we obtain three SPA pseudocodewords $\boldsymbol{\om}_{\Rm{C}}$, $\Rm{C}\in\{\Rm{R}, \Rm{G}, \Rm{B}\}$.
The output paths of Algorithm~\ref{alg:PCWD} given three SPA pseudocodewords $\boldsymbol{\om}_{\Rm{C}}$, $\Rm{C}\in\{\Rm{R}, \Rm{G}, \Rm{B}\}$, are listed in Table~\ref{table:SPA_LPPCWD_coc_eg_PCWD}.

There are two types of paths depending on whether the colors of the two endpoints are the same or not, i.e., $|c(S_i)| = 1$ or~$2$.
We discuss how to convert a path or a pair of paths into a vector based on the following two cases.
Here, we operate on supports of vectors instead of vectors.
\begin{enumerate}
\item ($|c(S_i)| = 1$) The path $P_i$ has two endpoints of the same color.
Such a path must be of even length $2\ell$ for some positive integer $\ell$ and it can be divided into $\ell$ pieces of length-$2$ paths, where each length-$2$ path can be converted into a vector consisting of two triangular faces, e.g., $P_2= \{e_{3,1}, e_{1,15}\}$ in Table~\ref{table:SPA_LPPCWD_coc_eg_PCWD} can be converted into a vector whose support is $V_1= \{v_{1,2,3}, v_{1,2,15}\}$ in Table~\ref{table:SPA_LPPCWD_coc_eg_PCWD_v}.
\item ($|c(S_i)| = 2$) The path $P_i$ has two endpoints of different colors.
Such a path must be of odd length $2\ell+1$ for some positive integer $\ell$ and it can be divided into a length-$(2\ell)$ path and a single edge $e'$, where the length-$(2\ell)$ path can be converted into a vector $\vv_i$ consisting of $2\ell$ triangular faces, e.g., $P_4 = \{e_{3,4}, e_{4,9}, e_{9,10}\}$ in Table~\ref{table:SPA_LPPCWD_coc_eg_PCWD} can be converted to be a vector of support $\{v_{3,4,5}, v_{4,5,9}\}$ and $e_{9,10}$.

There must exist another path $P_j$ such that $|c(S_i \cup S_j)| = 3$ and $|S_i \cup S_j|=3$ since an error at a triangular face introduces $3$ unsatisfied checks of $3$ different colors.
Suppose $P_j$ can be converted into a vector $\vv_j$ and a single edge $e''$.
The two paths $P_i$ and $P_j$ can be converted into a vector $\vv_i+\vv_j+\vv'$, where $\vv'$ is converted from $e'$ and $e''$.
For example, the two paths $P_4$ and $P_8$ in Table~\ref{table:SPA_LPPCWD_coc_eg_PCWD} can be converted to be a vector with support $V_5 $ in Table~\ref{table:SPA_LPPCWD_coc_eg_PCWD_v}, i.e., 
\begin{align}
&
\{e_{3,4}, e_{4,9}, e_{9,10}\} + \{e_{8,4}, e_{4,5}, e_{5,10}\} \\
&\mapsto 
\{v_{3,4,5}, v_{4,5,9}\}+e_{9,10} + \{v_{4,8,9}, v_{4,5,9}\}+e_{5,10}\\
&\mapsto \{v_{3,4,5}, v_{4,5,9}\} + \{v_{4,8,9}, v_{4,5,9}\} +\{v_{5,9,10}\}\\
&= \{v_{3,4,5}, v_{4,8,9}, v_{5,9,10}\} = V_5.
\end{align}
Note that $|\Supp(\vv')|$ can be $1$ or $3$, e.g.,
\begin{align}
\{e_{3,5},\ e_{4,5}\} &\mapsto \{v_{3,4,5}\} \text{ and }\\
\{e_{3,5},\ e_{5,10}\} &\mapsto \{v_{3,4,5}, v_{4,5,9}, v_{5,9,10}\}.
\end{align}
Note that after conversion, different paths or pairs of paths may result in the same vector, e.g.,
\begin{equation}
P_1+P_5\mapsto V_4 \text{ and } P_1+P_9 \mapsto V_4.
\end{equation}
\end{enumerate}

The paths in Table~\ref{table:SPA_LPPCWD_coc_eg_PCWD} are converted into ``generalized paths'', i.e., vectors in $\cC_X$, in Table~\ref{table:SPA_LPPCWD_coc_eg_PCWD_v} with the above mentioned procedure.
We formulate the problem of choosing some ``generalized paths'' from $\{\vv_i\}$ to obtain a vector with a matching syndrome as an LP, i.e.,
\begin{equation}
\hat{\mb{f}} \defeq \argmin \sum_i \lambda_i f_i,
\end{equation}
where $\lambda_i$ is the Hamming weight of $\vv_i$ and the constraints of the LP are as follows.
\begin{enumerate}
\item The variable $f_i\in\{0,1\}$ is indicating whether $\vv_i$ is chosen or not, which can be relaxed to $f_i\in[0,1]$.
\item Make sure that the syndrome $\vs$ is matched by satisfying 
$\sum_{j: i\in S'_j} f_j = 1\in\GF{2}\ \forall i \in \Supp(\vs)$, 
which is relaxed to
$\sum_{j: i\in S'_j} f_j = 1\ \forall i \in \Supp(\vs)$, i.e.,
\begin{alignat}{3}
& f_1+f_5 = 1,\ && f_3+f_5 = 1,\ && f_2+f_5=1,\\
& f_2+f_4 = 1,\ && f_3+f_4 = 1,\ && f_1+f_4=1.
\end{alignat}
\item At most one vector can be chosen when three vectors form a ``triangle'', i.e., $\sum_{i\in S} f_i \leq 1$ when $V_i \cap V_j \neq \emptyset$ for all $i\neq j \in S$ with $|S|=3$.
\end{enumerate}
The output of the LP is $\hat{\vf} = [0,0,0,1,1]$, corresponding to the vector whose support is associated with the gray triangles in Fig.~\ref{fig:color_eg1_path}.
\exampleend
\end{example}  
\section{Simulation Results}
\label{section4:simulation}
Fig.~\ref{fig:pSPA_coc} shows some simulation results of SPA+LPPCWD for color codes described in Algorithm~\ref{alg:SPA_LPPCWD_color}, where the maximum number of SPA iterations is 100.
As shown in Fig.~\ref{fig:pSPA_coc}, the performance of SPA+LPPCWD improves as the block length of color codes increases.
The performance of SPA+LPPCWD for color codes is similar to the performance of the decoders in~\cite{delfosse2014decoding,kubica2019efficient}, but the computational complexity of SPA+LPPCWD is much smaller than the computational complexity of the decoders in~\cite{delfosse2014decoding,kubica2019efficient}.
The threshold of SPA+LPPCWD, i.e., the crosspoint of lines, is close to $0.15$, where the rightmost points correspond to $p=0.15$.
Fig.~\ref{fig:pSPA_Wt_coc} shows the average/minimum weight of errors leading to decoding errors using SPA+LPPCWD for color codes.
We observe that the minimum weight of errors increases as the block length increases and the minimum weight of errors is at least $2L-1$ for each $L$. 
\begin{figure}
\begin{center}
\resizebox{1\columnwidth}{!}{\begin{tikzpicture}
\begin{loglogaxis}[
    xlabel={depolarizing probability $p$},
    ylabel={word error rate (WER)},
    legend pos=south east,
    grid=both,     
    width=0.5\textwidth
]

\addplot[color=blue,mark=*] coordinates { 
	(0.150000000000000,     0.566572237960340) 		
	(0.100000000000000,     0.362318840579710) 		
	(0.0630957344480193,   0.187090739008419)	 	
	(0.0398107170553497,   0.0879120879120879) 		
	(0.0251188643150958,   0.0374953130858643)	 	
	(0.0158489319246111,   0.0157257430413587) 		
	(0.0100000000000000,   0.00760225026607876) 		
	(0.0056250000000000,   0.00199962007218628) 		
	(0.0031640625000000,   0.000775990843308049) 	
    };

\addplot[color=red,mark=square*] coordinates { 
	(0.150000000000000,    0.668896321070234) 	    
	(0.100000000000000,    0.247524752475248) 	    
	(0.0630957344480193,  0.0571265352756355)	 	
	(0.0398107170553497,  0.00961446014806269) 		
	(0.0251188643150958,  0.00131675971768672)	    
	(0.0158489319246111,  0.000230287928997626) 	
	(0.0100000000000000,  3.43677215609682e-05) 	
	(0.0056250000000000,  3.96652946691998e-06) 	
    };

\addplot[color=orange,mark=pentagon*] coordinates { 
	(0.150000000000000,    0.621118012422360) 	    
	(0.100000000000000,    0.166163141993958) 	    
	(0.0630957344480193,  0.0204582651391162)	 	
	(0.0398107170553497,  0.00175201615599631)  		
	(0.0251188643150958,  0.000100931566820640)		
	(0.0158489319246111,  9.68618338793057e-06) 	
	(0.0100000000000000,  6.37796153284591e-07) 	
    };
   
\addplot[color=violet,mark=asterisk] coordinates { 
	(0.150000000000000,   0.785714285714286) 	    		
	(0.100000000000000,   0.139290407358739) 	    		
	(0.0630957344480193, 0.00676223965377333)	 	
	(0.0398107170553497,  0.000239808613541705) 	
	(0.0251188643150958,  9.05289550261582e-06)		
	(0.0158489319246111,  3.85632453130473e-07) 	
    };
    
\addplot[color=black,mark=x] coordinates { 
	(0.150000000000000,   0.692307692307692) 	    		
	(0.100000000000000,   0.0868945868945869) 	    
	(0.0630957344480193,  0.00208916114868454)	 	
	(0.0398107170553497,  3.65626599616373e-05) 	
	(0.0251188643150958,  3.31803855809657e-06)		
    };
    
\addplot[color=brown,mark=+] coordinates { 
	(0.150000000000000,   0.647058823529412) 	    		
	(0.100000000000000,   0.0655737704918033) 	    
	(0.0630957344480193,  0.000875213045280759)		
	(0.0398107170553497,  7.55863611969856e-06) 	
    };
    
\legend{$L = 1$, $L = 2$, $L = 3$, $L = 4$, $L = 5$, $L = 6$}
 
\end{loglogaxis}
\end{tikzpicture}}
\end{center}
\caption{Simulation results of SPA+LPPCWD for $\llbracket 18L^2,4,4L\rrbracket$ color codes.}
\label{fig:pSPA_coc}
\end{figure}
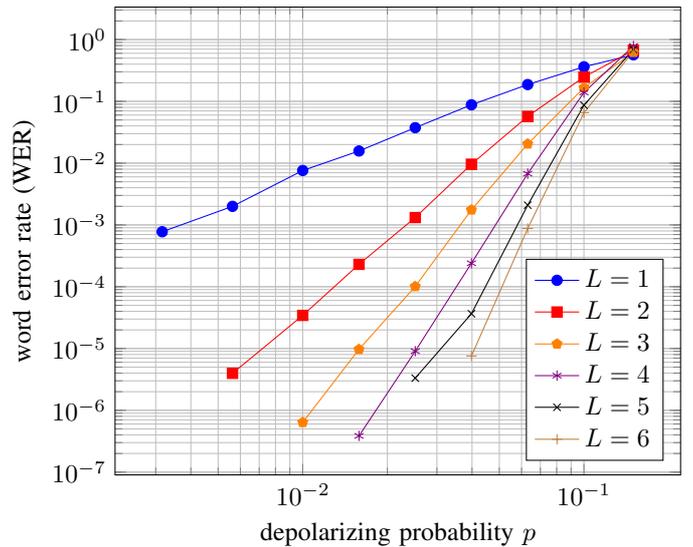
\begin{figure}
\begin{center}
\resizebox{1\columnwidth}{!}{\begin{tikzpicture}
\begin{semilogxaxis}[
    xlabel={depolarizing probability $p$},
    mark options={solid},
    xmin=0.0025, xmax=0.16,
    ymin=0, ymax=108.5,
    ylabel={average/minimum weight of errors},
    grid=both,
    width=0.5\textwidth,
    legend columns=2,
    legend style={
    legend pos=north west,
    /tikz/column 2/.style={
                column sep=5pt,},
	}
]

\addplot[color=blue,mark=*,error bars/.cd, y dir=both,y explicit] coordinates { 
	(0.150000000000000,     3.58500000000000) +- (1.23710433500105,1.23710433500105)		
	(0.100000000000000,     2.92000000000000) +- (0.904261268777479,0.904261268777479)		
	(0.0630957344480193,   2.66500000000000)	+- (0.752065831610089,0.752065831610089) 	
	(0.0398107170553497,   2.40500000000000) 	+- (0.680655167677755,0.680655167677755)    	
	(0.0251188643150958,   2.27000000000000)	+- (0.518114090770822,0.518114090770822) 	
	(0.0158489319246111,   2.14000000000000) 	+- (0.388787575540259,0.388787575540259)	
	(0.0100000000000000,   2.09000000000000) 	+- (0.303910857070336,0.303910857070336)	
	(0.0056250000000000,   2.07000000000000) 	+- (0.274731481513628,0.274731481513628)	
	(0.0031640625000000,   2.03500000000000) 	+- (0.184240939039320,0.184240939039320)	
    };
\addplot[dashed,color=blue,mark=*] coordinates { 
	(0.150000000000000,     2)	
	(0.100000000000000,     2)	
	(0.0630957344480193,   2) 	
	(0.0398107170553497,   2) 	
	(0.0251188643150958,   2) 	
	(0.0158489319246111,   2)	
	(0.0100000000000000,   2)	
	(0.0056250000000000,   2)	
	(0.0031640625000000,   2)	
    };
    
\addplot[color=red,mark=square*,error bars/.cd, y dir=both,y explicit] coordinates { 
	(0.150000000000000,    11.9250000000000) +- (2.60254514005620,2.60254514005620)	    	
	(0.100000000000000,    9.65500000000000) +- (2.11621521452116,2.11621521452116)	    	
	(0.0630957344480193,  7.78000000000000) +- (1.66295734631962,1.66295734631962)	 	
	(0.0398107170553497,  6.59500000000000) +- (1.48051837994901,1.48051837994901)	     
	(0.0251188643150958,  5.85500000000000) +- (1.17510824904982,1.17510824904982)        
	(0.0158489319246111,  5.10500000000000) +- (1.00449992558612,1.00449992558612)		
	(0.0100000000000000,  4.77000000000000) +- (0.806443896866961,0.806443896866961)		
	(0.0056250000000000,  4.42241379310345) +- (0.606499035003577,0.606499035003577)		
    };
\addplot[dashed,color=red,mark=square*] coordinates { 
	(0.150000000000000,     7)	
	(0.100000000000000,     6)	
	(0.0630957344480193,   4) 	
	(0.0398107170553497,   4) 	
	(0.0251188643150958,   4) 	
	(0.0158489319246111,   4)	
	(0.0100000000000000,   4)	
	(0.0056250000000000,   4)	
    };
    
\addplot[color=orange,mark=pentagon*,error bars/.cd, y dir=both,y explicit] coordinates { 
	(0.150000000000000,    26.2550000000000) +- (4.01003452409354,4.01003452409354)	     
	(0.100000000000000,    20.1636363636364) +- (3.09935557885347,3.09935557885347)	     
	(0.0630957344480193,  15.1750000000000) +- (2.87326526208975,2.87326526208975) 		
	(0.0398107170553497,  11.5954198473282) +- (2.16183193666212,2.16183193666212) 		
	(0.0251188643150958,  9.48437500000000) +- (1.98809201057690,1.98809201057690)    		
	(0.0158489319246111,  8.17948717948718) +- (1.98101064687292,1.98101064687292)		
	(0.0100000000000000,  6.88461538461539) +- (1.33647066789879,1.33647066789879)		
    };
\addplot[dashed,color=orange,mark=pentagon*] coordinates { 
	(0.150000000000000,     16)	
	(0.100000000000000,     13)	
	(0.0630957344480193,   8) 	
	(0.0398107170553497,   7) 	
	(0.0251188643150958,   5) 	
	(0.0158489319246111,   5)	
	(0.0100000000000000,   5)	
	(0.0056250000000000,   5)	
    };

\addplot[color=violet,mark=asterisk,error bars/.cd, y dir=both,y explicit] coordinates { 
	(0.150000000000000,   46)  +- (4.12887624513245, 4.12887624513245)	    	
	(0.100000000000000,   33.8773584905660)  +- (4.82000361625764, 4.82000361625764)	    
	(0.0630957344480193, 25.1700000000000)  +- (3.83249002581609, 3.83249002581609)	 	
	(0.0398107170553497,  18.8640000000000) +- (3.04891307972890, 3.04891307972890)		
	(0.0251188643150958,  14.1041666666667) +- (2.88974403146297, 2.88974403146297)	 	
	(0.0158489319246111,  9.87500000000000) +- (1.95940953204931, 1.95940953204931)		
    };
\addplot[dashed,color=violet,mark=asterisk] coordinates { 
	(0.150000000000000,     39)	    
	(0.100000000000000,     24)		
	(0.0630957344480193,   17) 	    
	(0.0398107170553497,   11) 		
	(0.0251188643150958,   9) 		
	(0.0158489319246111,   8)		
    };    
       
\addplot[color=black,mark=x,error bars/.cd, y dir=both,y explicit] coordinates { 
	(0.150000000000000,   69.0312500000000)  +- (5.70503102815910, 5.70503102815910)	    
	(0.100000000000000,   51.0655737704918)  +- (6.17879298293104, 6.17879298293104)	    
	(0.0630957344480193,  36.1186440677966) +- (4.78894108863727, 4.78894108863727) 		
	(0.0398107170553497,  26.8846153846154) +- (4.86889657377869, 4.86889657377869)	    
	(0.0251188643150958,  16.9375000000000) +- (4.17083125208073, 4.17083125208073) 		
    };
    
\addplot[dashed,color=black,mark=x] coordinates { 
	(0.150000000000000,     53)		
	(0.100000000000000,     37)		
	(0.0630957344480193,   28) 		
	(0.0398107170553497,   19) 		
	(0.0251188643150958,   11) 		
    };
    
\addplot[color=brown,mark=+,error bars/.cd, y dir=both,y explicit] coordinates { 
	(0.150000000000000,   100.090909090909)  +- (8.11764152301048, 8.11764152301048)	    
	(0.100000000000000,   74.4285714285714)  +- (5.64655628534755, 5.64655628534755)	    
	(0.0630957344480193,  49.4736842105263) +- (7.03458457316131, 7.03458457316131) 		
	(0.0398107170553497,  34.2500000000000) +- (7.88986691902975, 7.88986691902975)	    
    };
    
\addplot[dashed,color=brown,mark=+] coordinates { 
	(0.150000000000000,     85)	
	(0.100000000000000,     60)	
	(0.0630957344480193,   37) 	
	(0.0398107170553497,   25) 	
    };
\legend{$L = 1$ avg, $L = 1$ min, $L = 2$ avg, $L = 2$ min, $L = 3$ avg, $L = 3$ min, $L = 4$ avg, $L = 4$ min, $L = 5$ avg,  $L = 5$ min, $L = 6$ avg,  $L = 6$ min
}
 
\end{semilogxaxis}
\end{tikzpicture}}
\end{center}
\caption{Average/minimum weight of errors leading to decoding errors of SPA+LPPCWD for $\llbracket 18L^2,4,4L\rrbracket$ color codes.}
  \label{fig:pSPA_Wt_coc}
\end{figure}
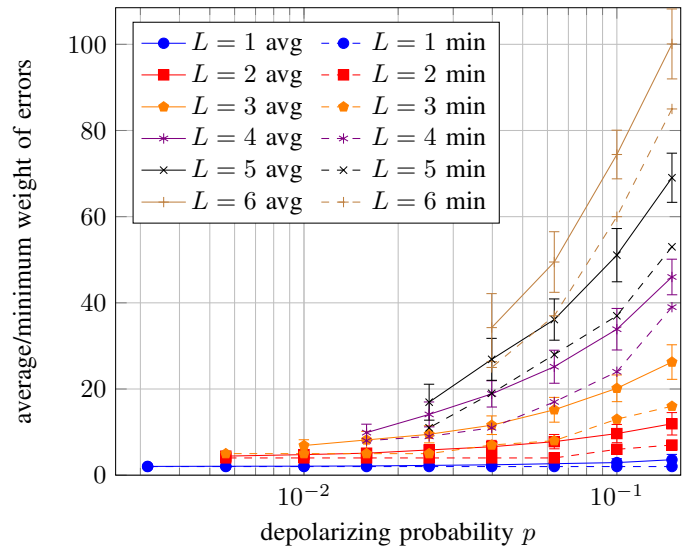

\bibliographystyle{IEEEtran}
\bibliography{ref}

\end{document}